\documentstyle[preprint,aps,prl,epsf]{revtex}   
\begin{document}
\title{{\bf Cooperativity and Contact Order in Protein Folding}}
\author{{\bf Marek Cieplak}}

\address{
Institute of Physics, Polish Academy of Sciences,
Al. Lotnik{\'o}w 32/46, 02-668 Warsaw, Poland \\}

\maketitle

\noindent

\vskip 40pt

\begin{abstract}
{The effects of cooperativity are studied within
Go-Lennard-Jones models of proteins by making the contact 
interactions dependent on the proximity to the native conformation. 
The kinetic universality classes 
are found to remain the same as in the absence of cooperativity.
For a fixed native geometry, small changes in the effective
contact map may affect the folding times in a chance way and to the
extent that is comparable to the shift in the folding times due to
cooperativity. The contact order controlls folding scenarios:
the average times necessary to bring pairs of amino acids into
their near native separations depend on the sequential distances
within the pairs. This dependence is largely monotonic,
regardless of the cooperativity,
and the dominant trend could be described by a single parameter
like the average contact order. However, it is the deviations
from the trend which are usually found to set the net folding times.
}
\end{abstract}

\vskip 40pt
PACS Numbers: 87.15.He, 87.15.Cc, 87.15.Aa

\newpage

There are many indications that properties of a protein, such as the
folding time, $t_{fold}$, depend on the topology of its
native state \cite{Shortle,Plaxco,Plaxco1}.
Furthermore, the number of possible native folds is known
to be limited \cite{Chothia}.
These two facts provide support for geometry based
modelling of proteins such as the tube-like \cite{tubes}
and Go-like approaches \cite{Goabe}.    
There are two geometrical parameters that have been
hypothesized as setting a scale for $t_{fold}$:
$N$ -- the number of amino acids (or the backbone
length of a protein) and $CO$ -- the relative contact
order parameter \cite{Plaxco}.  The latter is defined as an average
sequential distance between pairs of amino acids that
interact, or make a contact, and normalized by $N$.
A compilation of $t_{fold}$ that were measured at room temperature 
\cite{Plaxco,Plaxco1} suggested
a correlation with $CO$ but no dependence on $N$.
On the other hand, theoretical studies
yielded $t_{fold}$ growing with $N$ \cite{Thirumalai,biophysical}
and not depending on $CO$ \cite{biophysical}.
Is it the theories or the 
interpretation of the experimental data that are wrong?

Jewett, Pande and Plaxco \cite{Pande} have recently suggested
that the theories do not properly account for cooperativity effects,
i.e. for the fact that the strength of effective
amino acid interactions should depend on a conformation.
One of the reasons for such a dependence is that changes
in the conformation may lead to variations in the degree
of exposure to water molecules.
Specifically, Jewett et al. have considered a 27-mer
lattice Go model in which the contact energy, $\epsilon '$, is
related to the native contact energy, $\epsilon$,  through
\begin{equation}
\epsilon ' \;=\; \rho \epsilon\;\;,\;\;\;\;\;\; \rho=\frac{1}{(1-s)Q +s} \;\;,
\end{equation}
where $Q$ is the fraction of established native bonds and $s$
is a control parameter which introduces a
conformation dependence for $s > 1$.
The result of their simulations is that
$t_{fold}$ correlates with CO at the
57\% correlation level for $s$=3 as compared to 5\% for $s$=1 and
to 80\% reported in the experimental data. 
A related study of a similar model
by Kaya and Chan \cite{Kaya} indicates an even higher degree of
correlation induced by this kind of cooperativity.

What would this prescription for
cooperativity yield in off-lattice models of actual proteins?
In this Paper, we consider 14 $\alpha$-type (no native
$\beta$-sheets) and 16
$\beta$-type (no native helices) short proteins which
were studied in Ref. \cite{biophysical}.
The model is coarse-grained, i.e. it
involves only the $C^{\alpha}$ atoms, it is Go-like, i.e. the
potential is chosen so that the ground state agrees with the
experimentally determined native structure and it also favors the native
sense of chirality. We describe the contact interactions
by the Lennard-Jones  potential 
which is scaled by the energy
parameter $\epsilon '$ as in eq. (1). This parameter is identical
for all native contacts.
We take $s$=3 and adopt
an adiabatic way to smooth out any sudden changes in
$Q$ during the time evolution.

We find that the scaling of $t_{fold}$ with $N$ and 
the kind of the dependence on $CO$ do not change on switching
from $s=1$ to $s=3$, but the folding times  become, on average,
about 2.25 longer since the couplings provide a weaker pull
at the begining of the folding process.
In particular, we confirm existence of the
structure related kinetic universality classes \cite{biophysical}.
Even though the single $CO$ parameter itself does not
correlate with $t_{fold}$, the contact order, in a more general sense,
is important for the folding scenarios. One can characterize
the folding scenarios by plotting the average first times,
$t_c$, needed to establish specific native contacts as a
function of the corresponding sequence distance. We find that
the folding scenarios are governed by the
distance itself in a fairly monotonic way.
However, the control is
incomplete and often it is the out-of-trend deviations
that set the time to form the last contact, i.e that set $t_{fold}$.
Thus the local structures, like helices, do tend to form first
(in agreement with numerous experimental findings) but the way
the non-local structures form is not necessarily in agreement with
the contact order. Furthermore, we find that the effects
of cooperativity  may be less important than those of the precise
determination of the contacts considered native in the Go model.
Removing some contacts from the native set or declaring some
reasonable non-native contacts to be effectively native may
affect $t_{fold}$ more significantly than the tinkering with the strength
of the couplings. These kinds of adjustments in the contact map
physically correspond to considering sets of sequences which are
different and yet folding to nearly the same conformations, i.e.
belonging to the same native fold.

An important feature of our studies is that for each model
protein we determine the dependence of $t_{fold}$ 
on the temperature, $T$, and take $t_{fold}$ at the optimal
temperature, $T_{min}$, as the characteristic duration of folding
that is relevant for scaling studies.  When one considers
measurement or calculation at a fixed value of $T$
(such as the room temperature) then this $T$ can be in any 
distance away from the minimum of the typically U-shaped
dependence of $t_{fold}$ on $T$. Thus such,
essentially arbitrary, choices of $T$ may significantly
affect comparatory analysis of proteins. The arbitrariness is
removed if the kinetics are observed at $T_{min}$.
The experimental studies do not involve optimization of this kind.

When constructing the model, we follow references  \cite{Hoang}.
Briefly, the amino acids
are represented by particles of mass $m$ located at the
positions of the C$^{\alpha}$ atoms. They are tethered by a strong 
harmonic potential with a minimum at the peptide bond length. The
native structure of a protein is taken from the
Protein Data Bank \cite{PDB} and the interactions between the
amino acids are divided into native and non-native.
The distinction is based on the atomic representation
of the amino acids in the native state.
We check for overlaps between the atoms by associating spherical
volume to them. The assigned radii are 1.24 times
the van der Walls values and the multiplication factor accounts
for the softness of the potential \cite{Tsai}.   
The overlapping amino acids ($i$ and $j$)
are considered to be making contacts.
The resulting C$^{\alpha} - $ C$^{\alpha}$ 
contact separation ranges between 4.3 and
12.8 $\AA$. These pairs are
endowed with the Lennard-Jones potential 
in which the length
parameter $\sigma _{ij}$ is chosen  pair by pair
so that the native C$^{\alpha}$ -- 
C$^{\alpha}$ distance agrees with the minimum of the potential.
The non-native contacts are purely repulsive
(in the basic model) and  truncated at the distance of $2^{1/6}\sigma$
where $\sigma$ = 5$\AA$. During the time evolution, a
native contact is considered to be established when
the distance between the amino acids involved is less than
1.5 $\sigma _{ij}$.
The thermal fluctuations away from the native state are accounted
for through the Langevin noise
with the damping constant $\gamma$
of 2 $m/\tau$, where $\tau$ is $\sqrt{m \sigma ^2 /\epsilon}$.
This leads to negligible inertial effects 
\cite{biophysical}
but a more realistic account of the water environment
requires $\gamma$ to be at least an order of magnitude
larger. However,  $t_{fold}$ at $T_{min}$ has been found to be linear
in $\gamma$ \cite{Hoang} so the physical time scales can be accessed
by a simple rescaling.  

Figure 1 illustrates the way the model with the cooperativity
effect is defined by showing the time evolution of $\rho$.
The adiabatic way to incorporate  variations in $\rho$ is
as follows. The integration of the equations of motion is based
on the discretization of $\tau$
into 200 segments. With each time advancement by $\frac{1}{200}\tau$,
$Q$ that enters eq.1 becomes updated 
through $Q_{new}=0.99 Q_{old}+0.01 Q_{current}$ to eliminate rapid
jumps in $Q_{current}$. 
($Q_{current}$, the instantaneous value, is meant as $Q$
in Figure 1). 
The resulting $\rho$ is $\frac{1}{3}$ in the 
fully unfolded state as then $Q$ is zero. In the native state, 
$Q=\rho=1$. It is seen that at about 1/3 through of the
folding evolution, the variations in $\rho$ start mirroring
those in $Q$. The inset of Figure 1 shows that the initial reduction
in $\rho$, compared to the native value, results in a 
longer $t_{fold}$. However, the results for 
$t_{fold}$ at $s$=3 correlate
strongly with those at $s$=1. It should be noted
that the cooperativity effect is expected to enhance the thermodynamic
stability as it makes non-native local energy minima less stable 
relative to the native native state. 
On the other hand, we have found 
that the values of $T_{min}$ become lower (in the record cases
by 0.12). These two effects combined suggest that the energy landscape
gets sculpted in a way that enhances the folding funnel.

Figure 2 shows that cooperativity does not affect
the scaling curves. The largest value of $N$ considered here
is 154 and the smallest -- 35. In this range, it is hard to
distinguish between the power law and the exponential
dependencies, even though the correlation levels
for the $\alpha$ proteins favor the former slightly.
However,
there continues to be a support for existence of kinetic universality
classes that depend on the type of the secondary structure.
When the power law fits are used, 
the exponent for the $\alpha$-proteins
is about 1.7 and for the $\beta$-proteins -- about 3.2 (in the
mixed case it is about 2.5). The scaling trends seen in Figure 2
become disturbed, but still identifiable, when calculations are
performed not at $T_{min}$ but at a fixed $T$.

Cooperativity does not affect the dependence on $CO$ either,
as is shown in Figure 3. It is seen that a given value of
$CO$ may correspond to a big span of the values of $t_{fold}$
and there is no trend that can be demonstrated. Significant
variations in $t_{fold}$ can be obtained by staying with one
native geometry and adjusting the list of contacts
that are considered native in the dynamics. We focused on
three proteins: 1rpo, 1csp, and 1efn with the calculated
numbers of the contacts of 194 (N=61), 169 (N=67),
and 150 (N=57), respectively. We identified all non-native contacts
with the spatial C$^{\alpha}$--C$^{\alpha}$ range of less 
than 12 $\AA$ and considered
systems with 5, 10, 20, 30, and 40 such contacts, chosen
randomly, as providing additional active contacts. In addition, we
considered systems in which 7 long range native contacts were
removed from the native list. We studied variations of $t_{fold}$
with $CO$ within the three families of systems, each consisting
of 7 members.  The family of 1rpo shows a growing trend with $CO$.
This trend gets disturbed in the case of 1efn.
On the other hand, the variations around 1csp are chaotic.
It should be noted that the variations within the families
are not significant in the plots on the $N$-dependence, even
in the case of 1rpo. It should also be pointed out that our
data contain two pairs of proteins with identical
values of $N$ (2abd and 1imq in the $\alpha$ case and 1tit and 1ten
in the $\beta$ case) and in these pairs $t_{fold}$ is in fact longer
for the protein with the bigger $CO$.

In the Author's opinion, the experimental evidence, at the fixed
$T$, for the trends in $CO$ is not definitive. The inset of Figure 4
presents these data \cite{Plaxco1} 
separately for the $\alpha$- and $\beta$-proteins (the data
in the original paper are not plotted split into the three
structural classes: $\alpha$, $\beta$, and $\alpha - \beta$ ).
If one focuses just on the $\beta$ proteins then it emerges that four 
very different folding times correspond to almost the same $CO$. 
A similar point is demonstrated in the inset of Figure 5 which
presents the $\beta$-protein entries in the data compiled
by Galzitskaya et al. \cite{Finkelstein}.
The $\beta$-proteins form the crucial test case of the approach since they 
involve long range contacts. In the case of $\alpha$-proteins, $CO$ is
more a measure of the helical content, $h$, in the protein
than a measure of the sequence range which is short. The lower right
panel of Figure 6 shows that $h$, if non-zero,
is in fact anticorrelated with $CO$ to a fair degree (the
correlation coefficient of 0.74).

The folding scenarios, however, do depend on the contact order.
Not on its average value but on the full set of values. This is
illustrated in Figures 4 and 5 for the 1rpo ($\alpha$-type) and
1csp ($\beta$-type) proteins. The figures show $t_c$ as a function of the
sequence length $|j-i|$. 
In the case of the helical 1rpo, the
dependence is monotonic and as such it could be represented by
a single parameter, e.g. the relative (or average) contact order.
Note that there is no qualitative difference between the
$s$=3 and $s=1$ cases and the same goes to 1csp ($s=1$ not 
shown in Figure 5). Thus cooperativity does not  introduce
any new features in the folding scenarios other than general shifts.
Note that when 30 additional contacts that are consistent with the
native topology are introduced in 1rpo ($s$=3) then the shifts in the
data points are of the size that is comparable to
the very introduction of the cooperativity.
Thus cooperativity acts as if it was affecting the number of
effective contacts.

Figure 5 shows that the $\beta$-protein 1csp has a structured
form of the plot of $t_c$ vs. $|j-i|$. This kind of branched form 
is found in many other proteins both of the $\beta$-type, like in
1tit, and of the $\alpha$-type, like in 1f63, 1ycc, or 256b.
In 1csp, there are
many contacts of the same $|j-i|$ which are established at
different times. More importantly, the last to form are not 
the longest ranged contacts corresponding to $(i,j)$ = (1,62) and (1,64)
but the medium ranged (9,41), (8,43) and then (6,44), (6,45).
Adding the 20 contacts to 1csp shifts the pattern downward
but does not affect it in any fundamental manner. Note that
the formation time of the shortest ranged contacts is not affected
by the additional contacts. The acceleration of folding begins
in the second branch of contacts around $|j-i|$ of 11.

We have checked that the folding scenarios are not sensitive
to the details of the Go-modeling, such as the choice of the
contact potential (the 10-12 potential instead of the Lennard-Jones)
or to the addition of terms that depend on angular parameters,
such as the dihedral angles.
Furthermore, we have considered other variants
of the cooperativity effects. One of them is to replace $Q$ by
the ratio of the full potential energy of the system to its native value.
The results are qualitatively similar to those reported here
but the range of variations of the $\rho$ parameter during
the simulations is reduced, making it less effective,
compared to the contact based definition.

We conclude that incorporation of elements of cooperativity
in the couplings does not 
affect the folding process in the off-lattice Go models
in any qualitative manner other than making
the folding times longer despite the better sculpted folding funnel.
Notice that the parameter $Q$ does not generally couple to the
contact order. Thus it is puzzling why its incorporation into the
cooperativity effects appears to enhance the $CO$-dependence in the
$N$=27 lattice models. \cite{Pande,Kaya} 
Perhaps the lattice geometry itself imposes some sort of coupling
that emerges when considering systems of a fixed value of $N$.

Since the off-lattice Go models, with or without the cooperativity,
do not yield a correlation  of the folding times with $CO$ it is
interesting to ask whether some new quality arises if $CO$ is replaced
by the helical content parameter in the case of $\alpha$ and
$\alpha - \beta$ proteins. Figure 6 shows that this is not so.
The experimental data points compiled by Galzitskaya et al. 
\cite{Finkelstein} on two-state proteins containing helices
(the lower left panel) 
do anti-correlate with $h$ (the correlation coefficient is 0.68), 
though not as strongly as they
correlate with $CO$ (not shown; the correlation coefficient is 0.86). 
On the other hand, our model
calculations (the top two panels)
remain uncorrelated both with $h$ and with $CO$ even though
some tendency to grow with $h$ might be identified in the 
$\alpha - \beta$ case. We should reemphasize that the experimental
data do not pertain to the characteristic folding time that could
be uniquely associated with a protein. The characteristic time
must be obtained through an optimization process, i.e. it must
be measured at $T_{min}$. A room temperature based measurement
need not coincide with the conditions at $T_{min}$ and can
yield almost any value of $t_{fold}$, depending on the the precise
choice of the control parameters such as $T$ and pH.

We have considered a well controlled model of proteins and
demonstrated lack of dependence of the folding times on the relative
contact order parameter, irrespective of whether the cooperativity 
is taken into account or not. We have also demonstrated that, for 
a fixed geometry, or a fixed native fold,
one can get very different folding times depending on the sequence, i.e.
depending on what precisely constitutes the set of active contacts.
These findings do not mean that geometry of the native state is
irrelevant. Rather, they mean that the single relative contact order 
parameter
may be inappropriate to characterize it. It is the full native 
contact map that has a kinetic meaning.

MC appreciates fruitful discussions with T. X. Hoang and his help. 
The correspondence with K. W. Plaxco and especially his gift of 
reference \cite{Pande} before publication are also appreciated.
A. Sienkiewicz helped polish the manuscript.
This work was funded by KBN (grant 2 P03B 032 25).



\vspace*{2cm}
\centerline{FIGURE CAPTIONS}

\begin{description}
\item[Fig. 1.]
The temporal behavior of $Q$ (the thinner line) and $\rho$ 
(the thicker line) 
in an example of a folding trajectory
in the Go-like model of the 1csp protein with $s$=3.
The inset shows the median folding times determined with the
cooperativity factor ($s$=3) plotted vs. the folding times 
without any cooperativity effects ($s$=1). The hexagons are for the 
$\alpha$-proteins and the stars for the $\beta$-proteins. The straight
line corresponds to a 'conversion factor' of 2.25 ($\pm 0.2$).

\item[Fig. 2.]
The $N$-dependence of the median folding times, defined as the 'first
passage times', obtained based on at least 101 trajectories.
The top panels are 
for the $\alpha$-proteins and the bottom ones for the 
$\beta$-proteins. In the left-hand panels, the $N$ scale is
logarithmic and 
the corresponding power law exponents are indicated.
In, the right-hand panels, the $N$ scale is linear and 
the corresponding  correlation
lengths are indicated.
The PDB codes of
the $\alpha$-proteins studied are 1ce4, 1bba, 2pdd, 1bw6, 1rpo, 1hp8,
1ail, 2abd, 1imq, 1lmb, 1ycc, 1hrc, 256b, 1f63. The codes of the
$\beta$-proteins are: 1cbh, 1ixa, 1ed7, 1bq9, 1efn, 2cdx, 1csp, 2ait,
1bdo, 1tit, 1ten, 1wit, 1who, 6pcy, 1ksr, 4fgf.
The correlation levels of the power law (exponential)
fits are 0.978 and 0.956 (0.960 and 0.971)
for the $\alpha$ and $\beta$ proteins respectively. 

\item[Fig. 3.]
The dependence of $t_{fold}$ on the relative contact order
for the $\alpha$- (triangles)
and $\beta$-proteins (circles).
The dotted line separates the values
of CO that were found for the $\alpha$- from those found
for the $\beta$-proteins. The filled symbols correspond to three
selected proteins 1rpo (circles), 1csp (squares), and 1efn (stars)
in which extra contacts were added or some contacts were subtracted.
The families of such systems are connected by lines.
The results corresponding to the true native contact maps
are shown by the larger symbols.
Within each family, $t_{fold}$
at an individually determined optimal temperature
is displayed. If a fixed temperature is used instead (the one which is
optimal for the true native contact map) the plots would look similar
in character. The folding times of 7019 $\tau$ for 1f63 ($CO$=0.1291),
5024 $\tau$ for 6pcy ($CO$=0.2448), and 19600 $\tau$ 
for 4fgf ($CO$=0.1873) are
beyond the vertical scale of this figure.

\item[Fig. 4.]
The folding scenarios for the 1rpo protein
as described by the average time to form
a contact corresponding to the sequence length $|j-i|$. The squares
correspond to the standard Lennard-Jones Go-like model whereas the
stars are for the systems with the couplings modified by the
cooperativity effect. The small dots indicate data obtained when 
30 non-native contacts are added randomly (with the condition that
the contacts formed are shorter than 12 $\AA$
in the physical space) and the cooperativity
factor $\rho$ corresponds to $s$=3.
The data are based on 200 trajectories at $T_{min}$.
The inset shows compilation of the
experimental results, based on the data from \cite{Plaxco1}.
The relative contact order CO$_P$ is calculated somewhat differently
than CO in that it involves non-hydrogen atoms in a distance less
than a cutoff value of 6 $\AA$ as discussed further in ref. \cite{biophysical}.
CO involves only the $C^{\alpha}$ atoms but 
existence of a contact is based on the atomic overlap.

\item[Fig. 5.]
The folding scenario for the 1csp protein with the cooperativity
effect included. The small dots indicate data obtained when 20 
non-native contacts are added randomly.
The inset shows the two-state protein data
compiled by Galzitskaya et al. \cite{Finkelstein}
and plotted vs. the relative contact order parameter, $CO_G$, as 
calculated by them -- usually it coincides with $CO_P$.

\item[Fig. 6]
The dependence on the helical content parameter $h$ defined
as the ratio of the number of amino acids that belong to $\alpha$-helices
to the total number of amino acids in a protein.
The top two panels shows results of the model calculations for $s$=1
as obtained in
Ref. \cite{biophysical} for the $\alpha - \beta$ and $\alpha$
proteins respectively. The proteins considered are those listed in this
reference. The bottom panel on the right shows the dependence of
$CO$ on $h$ for the same proteins. The bottom panel on the left plots
the experimental data points \cite{Finkelstein} for a different set of
proteins (there is a substantial overlap with the set considered in the
simulations).

\end{description}

\begin{figure}
\epsfxsize=6in
\centerline{\epsffile{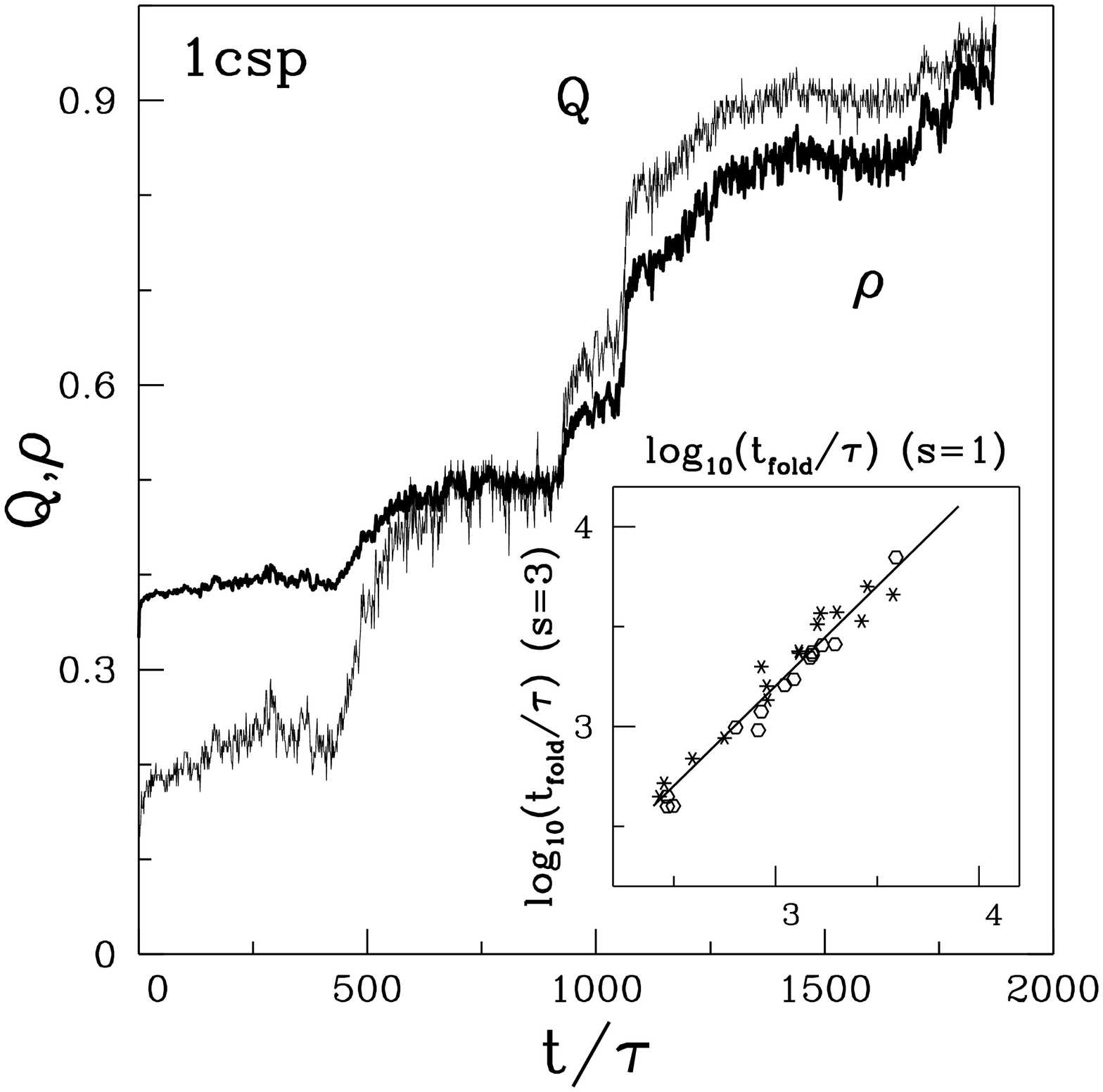}}
\caption{ }
\end{figure}

\begin{figure}
\epsfxsize=7in
\centerline{\epsffile{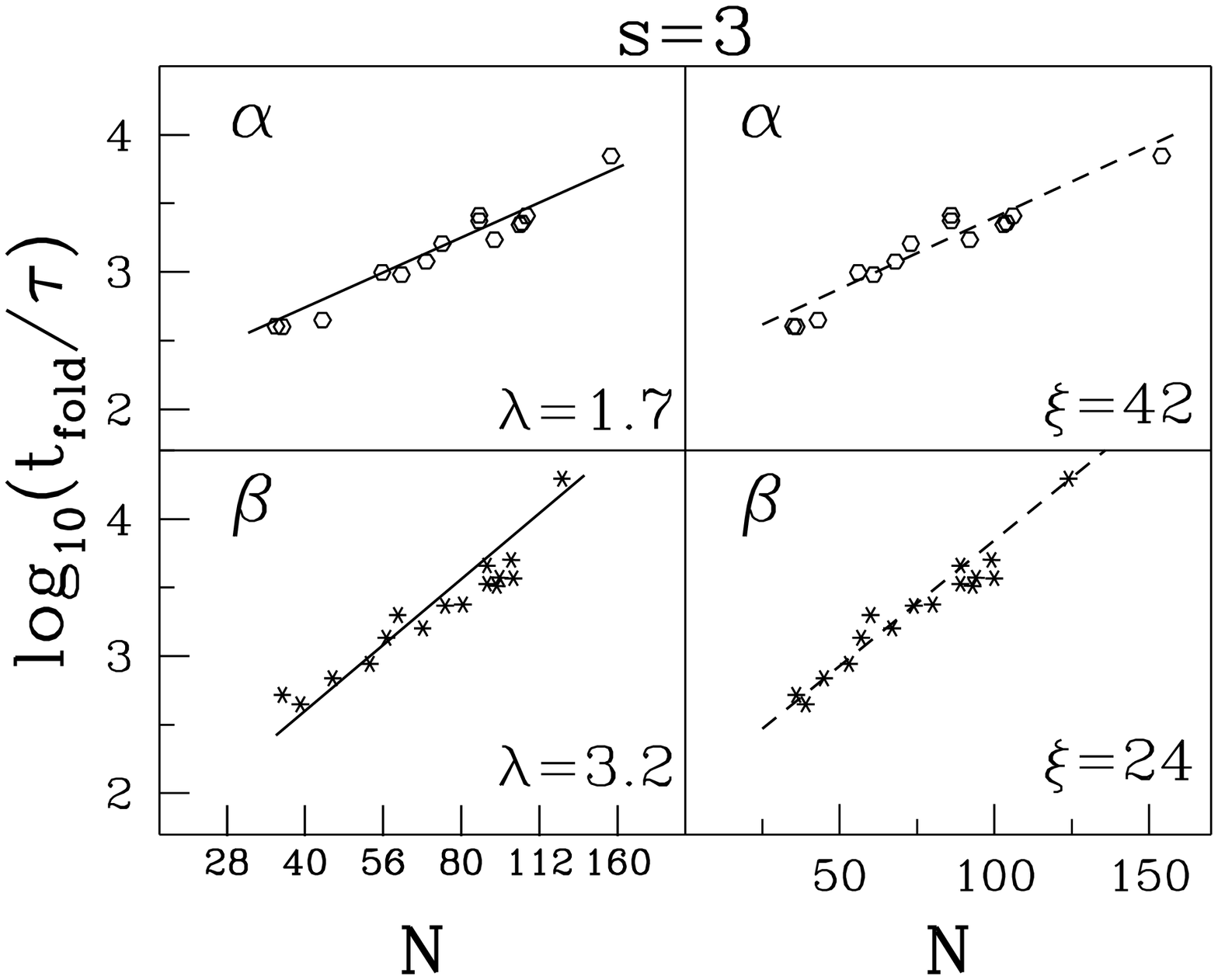}}
\vspace*{3cm}
\caption{ }
\end{figure}

\begin{figure}
\epsfxsize=7in
\centerline{\epsffile{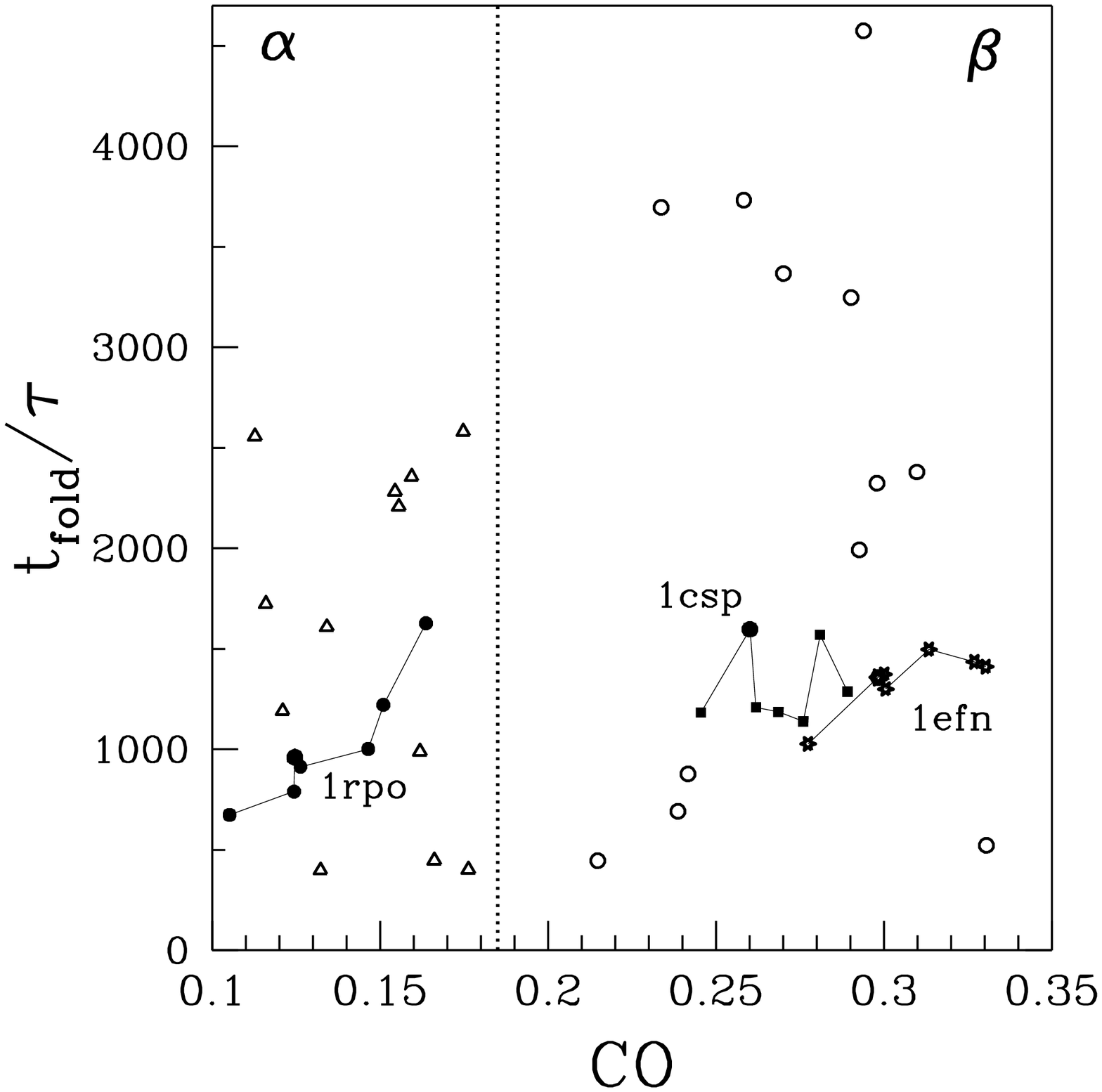}}
\vspace*{3cm}
\caption{ }
\end{figure}

\begin{figure}
\epsfxsize=7in
\centerline{\epsffile{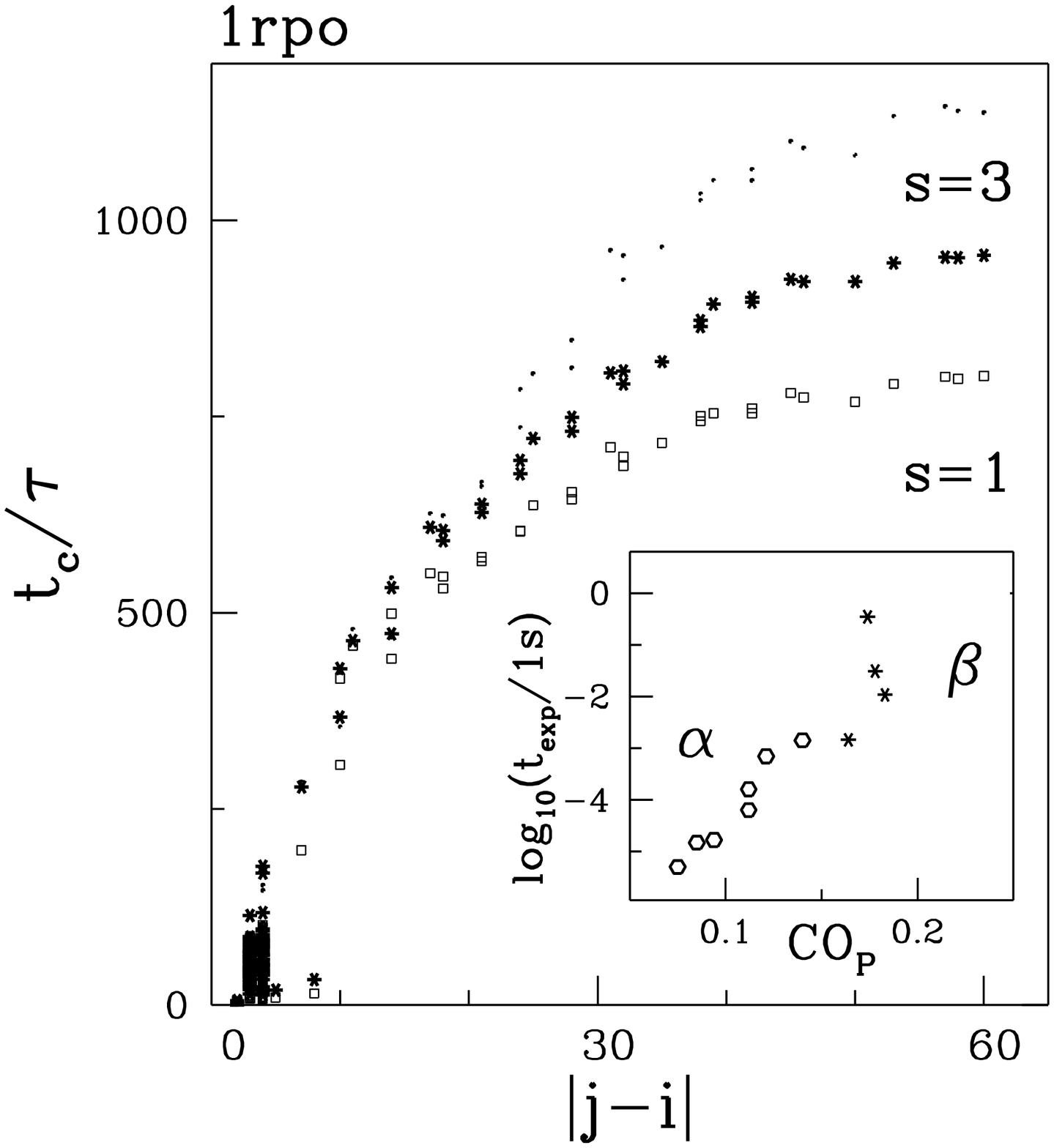}}
\vspace*{3cm}
\caption{ }
\end{figure}

\begin{figure}
\epsfxsize=7in
\centerline{\epsffile{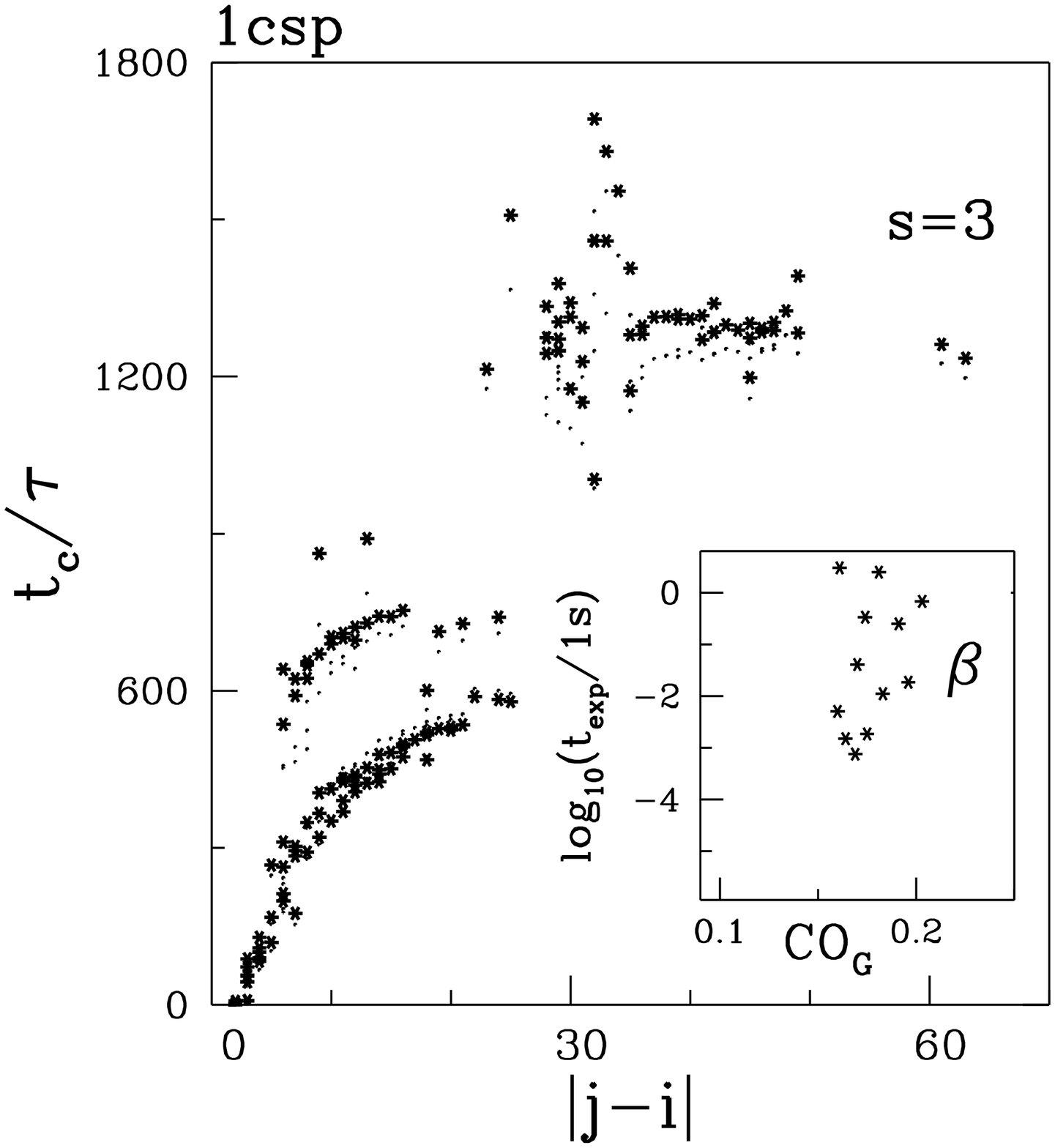}}
\vspace*{3cm}
\caption{ }
\end{figure}

\begin{figure}
\epsfxsize=7in
\centerline{\epsffile{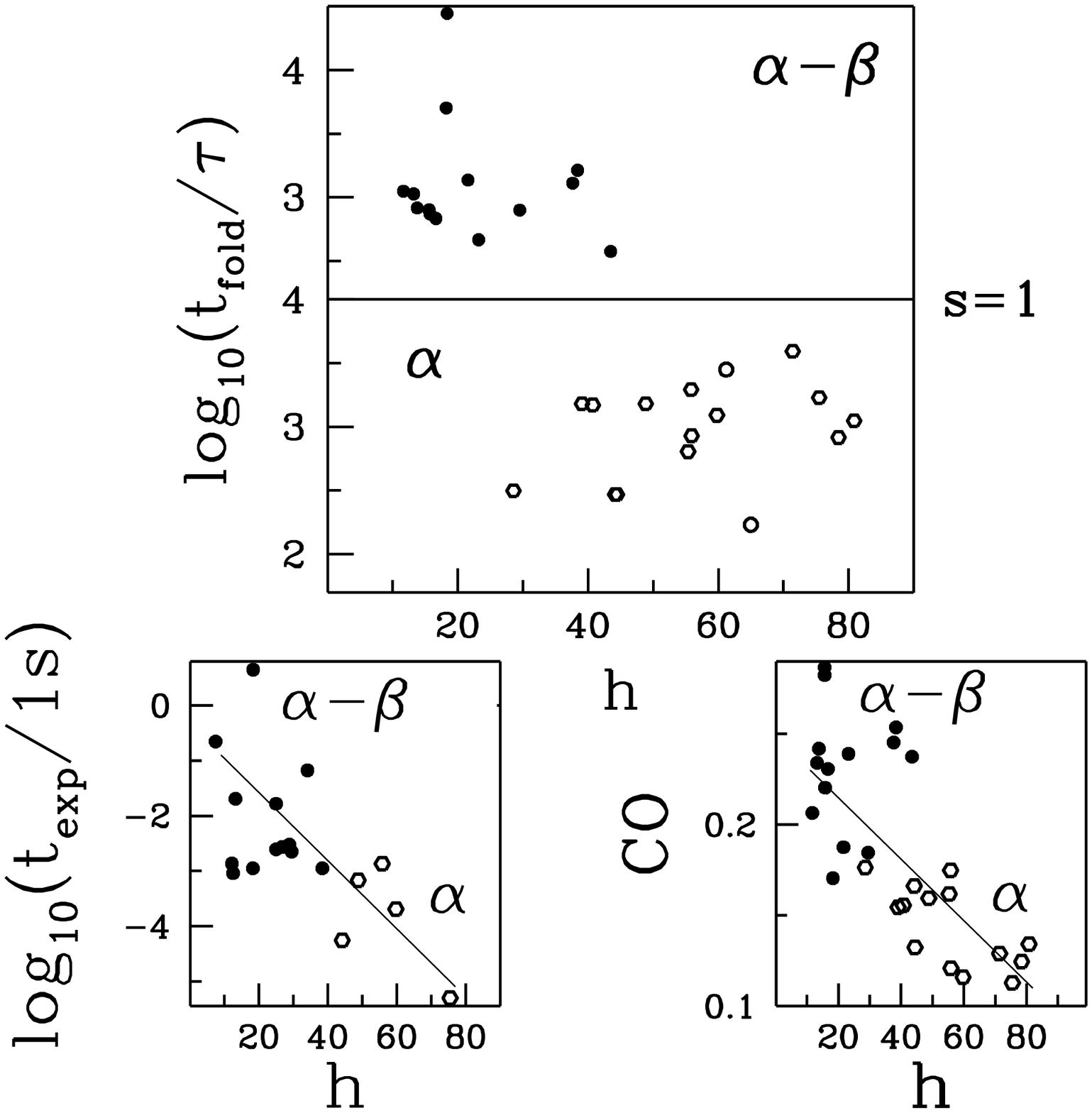}}
\vspace*{3cm}
\caption{ }
\end{figure}

\end{document}